\documentclass[aps, a4paper,superscriptaddress, nofootinbib,12pt]{revtex4-1}
\usepackage{amsmath,amssymb, bm, natbib}
\usepackage{dsfont}
\usepackage{epsfig}
\usepackage{graphicx}
\usepackage{slashed}
\usepackage{color}

\usepackage{accents}
\usepackage{mathrsfs}
\usepackage[center,footnotesize,hang]{subfigure}
\usepackage{bigstrut}

\begin{document}

\title{Dirac or Inverse Seesaw Neutrino Masses with\\ $B-L$ Gauge Symmetry and 
$S_3$ Flavour Symmetry}

\author{Ernest Ma}
\email{ma@phyun8.ucr.edu}
\affiliation{Department of Physics and Astronomy, University of California, \\
Riverside, California 92521, USA}
\author{Rahul Srivastava}
\email{rahuls@imsc.res.in}
\affiliation{The Institute of Mathematical Sciences, Chennai 600 113, India}


\begin{abstract}
   \vspace{1cm}
   
  \begin{center}
\textbf{ Abstract} 
\end{center}
Many studies have been made on extensions of the standard model with $B-L$
gauge symmetry. The addition of three singlet (right-handed) neutrinos 
renders it anomaly-free. It has always been assumed that the spontaneous 
breaking of $B-L$ is accomplished by a singlet scalar field carrying two 
units of $B-L$ charge. This results in a very natural implementation of the 
Majorana seesaw mechanism for neutrinos. However, there exists in fact 
another simple anomaly-free solution which allows Dirac or inverse seesaw 
neutrino masses. We show for the first time these new possibilities and 
discuss an application to neutrino mixing with $S_3$ flavour symmetry.

\end{abstract}

\maketitle


\section{INTRODUCTION}
\label{sec1}


 The standard model (SM) of quarks and leptons is free of gauge anomalies. 
If it is extended to include an extra $U (1)$ gauge group, the conditions 
for the absence of gauge anomalies impose nontrivial constraints. 
Nevertheless, interesting discoveries of such possibilities have been made. 
One example \cite{Ma:2001kg} adds a lepton triplet per family to the SM. 
Another \cite{Ma:2002tc} adds a number of new superfields to the minimal 
supersymmetric standard model (MSSM) of three families. On the other hand, 
the most studied extension is that of $U (1)_{B-L}$ .  Using
the particle content of the standard model, all triangle gauge anomalies 
are zero except for

\begin{eqnarray}
\sum U(1)^3_{B-L} = -3 
\label{sum}
\end{eqnarray}
 This is easily solved with the addition of three (right-handed) singlet 
neutrinos, which contribute 
 $-3(-1)^3 = +3$. Note that the gauge-gravitational anomaly is also zero because
$-3(-1) = +3$. Numerous studies have been made regarding this model of 
$U (1)_{B-L}$ .  \\

In this paper, we point out that there is another simple choice for three 
families \cite{Montero:2007cd, Machado:2010ui,Machado:2013oza}. Let $\nu_{Ri} \sim n_i$ 
under $B-L$, then $n_{1,2,3} = (+5,-4,-4)$ yields

\begin{eqnarray}
  -(+5)^3 -(-4)^3  -(-4)^3    = +3
 \label{rn}      
\end{eqnarray}
 as well. In fact, since $- (5) -(-4) - (-4)  = +3$, the mixed 
gauge-gravitational anomaly is also zero.  Now the standard-model Higgs 
doublet $(\phi^+ , \phi^0 )^{T}$ does not connect $\nu_L$ with $\nu_R$
and there is no neutrino mass. Consider then three heavy Dirac singlet 
fermions $N_{L,R}$ , all transforming as $-1$ under $B-L$. They do not 
change the anomaly conditions, but now in the case of 
$\nu_{R2}$ and $\nu_{R3}$, $(\bar{\nu}_L, \bar{N}_L)$ is linked 
to $(\nu_R,N_R)$ through the $2 \times 2$ mass matrix  as follows

  \begin{equation}
M_{\nu,N} = \left( 
\begin{array}{cc}
0     & m_0  \\
m_3   & M  
\end{array}
\right)
 \label{dseesaw}
\end{equation}
where $m_0$ comes from $\langle \phi^0 \rangle$ and $m_3$ comes from 
$\langle \chi_3\rangle$ , assuming the Yukawa coupling 
$\bar{N}_L \nu_R \chi_3$, i.e. $\chi_3 \sim +3$ under $B - L$. Now, the 
invariant mass $M$ is naturally large, so the Dirac seesaw \cite{Roy:1983be} 
yields a small neutrino mass $m_3 m_0 /M$ . In the conventional $U (1)_{B-L}$ 
model, $\chi_2 \sim +2$ under $B-L$, is chosen to break the gauge symmetry, so that 
$\nu_R$ gets a Majorana mass and lepton number $L$ is broken to $(-1)^L$. 
Here, $\chi_3 \sim +3$ means that $L$ remains a conserved global symmetry, 
with $\nu_{L,R}$ and $N_{L,R}$ all having $L = 1$. Since $\nu_{R1} \sim +5$ 
does not connect with $\nu_L$ or $N_L$ directly, there is one massless 
neutrino in this case. However, the dimension-five operator
$\bar{N}_L \nu_{R3}\chi^*_3 \chi^*_3/\Lambda$ is allowed by $U (1)_{B-L}$ and 
would give it a very small Dirac mass. Alternatively, a second scalar 
$\chi_6 \sim 6$ may be added. \\

The above assignment of $B-L = +3$ for the scalar field $\chi_3$; is the 
new idea of this paper, which was not recognized anywhere before. 
Because the pairing of two neutral fermions of the same chirality in 
this model always results in a nonzero $B-L$ charge not divisible by 3 units, 
it is impossible to construct an operator of any dimension for a 
Majorana mass term which violates $B-L$.  Hence the statement that $B-L$ is 
reduced to a conserved global $U(1)$ symmetry is definitely robust.\\

 Another possible outcome of the above scenario is obtained with the choice 
of two complex scalar fields $\chi_2 \sim 2$ and $\chi_6 \sim 6$ under 
$U (1)_{B-L}$. In this case, $\nu_L$ is not connected to
$\nu_{R1,R2} \sim -4$. It is connected however to $N_{L,R}$ through the mass 
matrix spanning $(\bar{\nu}_L, N_R, \bar{N}_L)$ as follows:

  \begin{equation}
M_{\nu,N} = \left( 
\begin{array}{ccc}
0     & m_0     & 0 \\
m_0   & m'_2   & M \\
0     & M       & m_2
\end{array}
\right)
 \label{invseesaw}
\end{equation}
where $m_2$ and $m'_2$ come from $\chi_2$ . This is then an inverse 
seesaw \cite{Wyler:1982dd, Mohapatra:1986bd, Ma:1987zm}, i.e. 
$m_\nu \simeq m^2_0 m_2/M^2$. In the case of $\nu_{R3} \sim +5$, the 
corresponding mass matrix spanning $( \bar{\nu}_L, N_R, \bar{N}_L, \nu_{R3})$ 
is given by

 \begin{equation}
M_{\nu,N} = \left( 
\begin{array}{cccc}
0     & m_0     & 0    & 0    \\
m_0   & m'_2    & M    & 0    \\
0     & M       & m_2  & m_6   \\
0     & 0       & m_6  & 0
\end{array}
\right)
 \label{lopseesaw}
\end{equation}
where $m_6$ comes from $\chi_6$ . Thus $\nu_{R3}$ also gets an inverse seesaw 
mass $\simeq m^2_6 m'_2 / M^2$. Note that Eq. (\ref{lopseesaw}) is the 
$4 \times 4$ analog of the $3 \times 3$ lopsided seesaw discussed 
in \cite{Ma:2009du}.     \\

In this scheme, lepton number $L$ is broken to $(-1)^L$ for $\nu_L$ and 
$\nu_{R3}$, and $N_{1,2,3}$ are heavy pseudo-Dirac fermions. However, 
$\nu_{R1,R2}$ remain massless and may only be produced in pairs. They are 
thus good dark-matter candidates if they become massive. This may be
accomplished with a third scalar $\chi_8 \sim 8$.


\section{The $S_3$ Flavour Symmetry: $\mu-\tau$ sector}
\label{sec2}


In this section we expand upon the first scenario discussed in the previous 
section, using two singlet scalar fields transforming as $+3$ under the 
$B-L$ symmetry, in analogy to having $\nu_{R2}$ and $\nu_{R3}$ transforming as 
$-4$.  This allows us to use the doublet representations of the non-Abelian 
discrete symmetry group $S_3$ to understand the leptonic family structure 
as indicated by present neutrino oscillation data and other experimental 
constraints.  Our application of $S_3$ will be different from those of 
Ref.~\cite{Machado:2010ui,Machado:2013oza} since we have the three $N_{L,R}$ fields 
in addition to $\nu_{R1,R2,R3}$. 
We will first work out the details of our model for the simpler 
case of $\mu-\tau$ sector and show how one can use $S_3$
symmetry to obtain maximal mixing.  Then we will discuss the implications of 
imposing $S_3$ symmetry to the full $3\times 3$ mixing matrix. However, 
before presenting our model let us briefly discuss the $S_3$ symmetry group.\\

The $S_3$ group is the smallest non-Abelian discreet symmetry group and is 
the group of the permutation of three objects. It consists of six elements 
and is also isomorphic to the symmetry group of the equilateral triangle. 
It admits three irreducible representations
$1$, $1'$ and $2$ with the tensor product rules

\begin{eqnarray}
  1 \times 1' = 1', \quad 1' \times 1' = 1, \quad 2 \times 1 = 2, \quad 
2\times 1' = 2, \quad 2\times 2 = 1 + 1' +2
 \label{prule}
\end{eqnarray}

In this work, following the earlier works \cite{Chen:2004rr,Ma:2013zca}, we 
choose to work with the complex representation of the $S_3$ group. In the 
complex representation, if

 \begin{eqnarray}
 \left( \begin{array}{c}
\phi_1       \\
\phi_2     
\end{array} \right)\, , \quad
 \left( \begin{array}{c}
\psi_1       \\
\psi_2     
\end{array} \right)\,  \in \textbf{2}
\quad \Rightarrow \quad
\left( \begin{array}{c}
\phi^\dagger_2       \\
\phi^\dagger_1     
\end{array} \right)\, , \quad
 \left( \begin{array}{c}
\psi^\dagger_2        \\
\psi^\dagger_1     
\end{array} \right)\,  \in \textbf{2}
 \label{s3-2d-rep}
\end{eqnarray}
such that

 \begin{eqnarray}
  \phi_1 \psi_2 +  \phi_2 \psi_1 \, , \quad   \phi^\dagger_2 \psi_2 +  
\phi^\dagger_1 \psi_1  \quad \in \textbf{1}  \nonumber \\
  \phi_1 \psi_2 -  \phi_2 \psi_1 \, , \quad   \phi^\dagger_2 \psi_2 -  
\phi^\dagger_1 \psi_1  \quad \in \textbf{1}' \nonumber \\
  \left( \begin{array}{c}
\phi_2 \psi_2      \\
\phi_1 \psi_1    
\end{array} \right)\, , \quad
 \left( \begin{array}{c}
\phi^\dagger_1  \psi_2       \\
\phi^\dagger_2   \psi_1    
\end{array} \right)\,  \in \textbf{2}
   \label{s3tpro}
 \end{eqnarray}

With this brief summary of $S_3$ group and its irreducible representations, 
we now move on to constructing an $S_3$ invariant $\mu-\tau$ sector. In later 
section we will generalize the discussion of this section to the full 
$3\times 3$ mixing matrix.


\subsection{The $S_3$ invariant $\mu-\tau$ sector}
\label{sub2-1}


 We denote the left handed lepton doublets by $L^\alpha = \left( \nu^\alpha_L , 
l^\alpha_L  \right)^{\rm{T}}$ where $\alpha = \mu, \tau$; the right handed 
charged leptons are denoted as $\mu_R, \tau_R$ and the right handed neutrinos 
as $\nu^{\mu}_R, \nu^{\tau}_R$. Also, let us denote the 
 heavy singlet fermions as $N^i_{L,R}$; $i=2,3$. The ``Standard Model like'' 
scalar doublet is denoted by $\Phi = \left( \phi^+ , \phi^0  \right)^{\rm{T}}$ 
and the singlet scalars are denoted by $\chi_{2,3} $.   \\

Let the  $B-L$ charge and $S_3$ assignment of the above fields be as shown 
in Table \ref{tab1}.   

\begin{table}[h]
\begin{center}
\begin{tabular}{c c c || c c c}
  \hline \hline
  Fields                      \hspace{2cm}                    & $B-L$                 \hspace{2cm}                                     &  $S_3$       \hspace{2cm}              & 
  Fields                      \hspace{2cm}                    & $B-L$                 \hspace{2cm}                                     &  $S_3$                                 \\
  \hline \hline
  $L^\mu$                     \hspace{2cm}                    & $-1$                   \hspace{2cm}                                    &  $1'$            \hspace{2cm}           &   
  $L^\tau$                    \hspace{2cm}                    & $-1$                    \hspace{2cm}                                   &  $1$                                    \\
  $\mu_R$                     \hspace{2cm}                    & $-1$                   \hspace{2cm}                                    &  $1'$            \hspace{2cm}           &   
  $\tau_R$                    \hspace{2cm}                    & $-1$                    \hspace{2cm}                                   &  $1$                                    \\
  $N^2_L $                     \hspace{2cm}                   & $-1$                     \hspace{2cm}                                  &  $1'$               \hspace{2cm}        & 
  $N^3_L $                      \hspace{2cm}                  & $-1$                    \hspace{2cm}                                   &  $1$                                    \\
  $N^2_R $                      \hspace{2cm}                  & $-1$                     \hspace{2cm}                                  &  $1'$               \hspace{2cm}         & 
  $N^3_R $                        \hspace{2cm}                & $-1$                     \hspace{2cm}                                  &  $1$                                     \\
   $\Phi$              \hspace{2cm}                           & $0$                      \hspace{2cm}                                  & $1$           \hspace{2cm}               &
   & & \\  
       $ \left( \begin{array}{c}
  \nu^\mu_R   \\ \nu^\tau_R   
  \end{array} \right)$               \hspace{2cm}               & $ -4$                     \hspace{2cm}                                 &  $2$            \hspace{2cm}           &    
  \vspace{0.5cm}
  $ \left( \begin{array}{c}
  \chi_2   \\ \chi_3                                                           
  \end{array} \right)$               \hspace{2cm}               & $ 3$                         \hspace{2cm}                             &  $2$                                    \\
  \hline
  \end{tabular}
\end{center}
\caption{The $B-L$ and $S_3$ charge assignment for the fields.}
  \label{tab1}
\end{table}    

The $S_3$ and $B-L$ invariant Yukawa $\mathcal{L}^{(2)}_Y$ interaction can be 
written as 

\begin{eqnarray}
  \mathcal{L}^{(2)}_Y & = &    \mathcal{L}^{(2)}_{L^\alpha l_R} }  \, + \,   \mathcal{L}^{(2)}_{L^\alpha N_R}   \, + \,  \mathcal{L}^{(2)}_{N_L N_R}      \, + \,  \mathcal{L}^{(2)}_{N_L \nu_R
 \label{yuk2}
\end{eqnarray}

 where 
 
\begin{eqnarray}
\mathcal{L}^{(2)}_{L^\alpha l_R}  & = & y_\mu \, \bar{L}^\mu \Phi \mu_R    \, + \,   y_\tau \,  \bar{L}^\tau \Phi \tau_R           \nonumber \\
\mathcal{L}^{(2)}_{L^\alpha N_R}  & = & g_2 \, \bar{L}^\mu \hat{\Phi}^* N^2_R  \, + \,  g_3  \, \bar{L}^\tau \hat{\Phi}^* N^3_R     \nonumber \\
\mathcal{L}^{(2)}_{N_L N_R}       & = &  M_2 \, \bar{N}^2_L N^2_R  \, + \,  M_3 \, \bar{N}^3_L N^3_R                               \nonumber \\
\mathcal{L}^{(2)}_{N_L \nu_R}     & = &  f_2 \, \bar{N}^2_L \, \otimes \, \left[  \left( \begin{array}{c}  \nu^\mu_R   \\ \nu^\tau_R  \end{array} \right) \, \otimes \, \left( \begin{array}{c}  \chi_2   \\ \chi_3   \end{array} \right) \right]_{1'}  
\, + \,  f_3 \, \bar{N}^3_L \, \otimes \, \left[  \left( \begin{array}{c}  \nu^\mu_R   \\ \nu^\tau_R  \end{array} \right) \, \otimes \, \left( \begin{array}{c}  \chi_2   \\ \chi_3   \end{array} \right) \right]_{1}                                                                    
\label{cly2}
\end{eqnarray}

In writing (\ref{cly2}), we used the notation $\hat{\Phi}^* = i \tau_2 \Phi^* = (\phi^0, \phi^-)^{\rm{T}}$. Moreover, $y_\mu$, $y_\tau$ are the Yukawa couplings of the charged lepton sector whereas  $f_i$, $g_i$ and $M_i$ 
denote the dimensionless coupling constants between the leptons and the heavy fermions.

After symmetry breaking the scalar fields get vacuum expectation values (VEVs) $\langle \phi^0 \rangle = v$, $\langle \chi_i \rangle = u_i$; $i = 2,3$.  Then the mass matrix relevant to charged leptons is given by 

 \begin{eqnarray}
   \mathcal{M}_l =   v \, \left( \begin{array}{c c} 
                      y_\mu     & 0      \\ 
                      0        &  y_\tau
                      \end{array} \right) 
  \label{lm2}
 \end{eqnarray}

Also, the $4\times 4$ mass matrix spanning $(\bar{\nu}^\mu_L, \bar{\nu}^\tau_L, \bar{N}^2_L, \bar{N}^3_L)$ and $(\nu^\mu_R, \nu^\tau_R, N^2_R, N^3_R)^{\rm{T}}$ of neutrinos and the heavy fermions is given by

\begin{eqnarray}
   \mathcal{M}_{\nu,N} =    \, \left( \begin{array}{c c  c c} 
                      0                &   0                 &  g_2 v^*             &   0                       \\ 
                      0                &   0                 &  0                   &    g_3  v^*                \\ 
                      f_2 u_3    &   -f_2 u_2    &  M_2                 &   0                        \\ 
                      f_3 u_3    &   f_3 u_2     &  0                   &  M_3 
                      \end{array} \right) 
  \label{lnuN2}
 \end{eqnarray}

 As remarked earlier, the mass terms $M_i$ between the heavy fermions can be 
naturally large, so we can block diagonalize (\ref{lnuN2}) assuming that 
$f_i, g_i << M_i$. The block diagonalized mass matrix of light neutrinos is 
given by
 
 \begin{eqnarray}
   \mathcal{M}_\nu & = &   m^{(2)}_{N_L \nu_R} \frac{1}{M^{(2)}_{N_L N_R}} m^{(2)}_{L^\alpha N_R}  \nonumber \\
                   & = &      v^* \, \left( \begin{array}{c c} 
                              \frac{f_2 g_2 }{M_2}  u_3        & - \frac{f_2 g_3 }{M_3}  u_2      \\ 
                              \frac{f_3 g_2 }{M_2}  u_3        &  \frac{f_3 g_3 }{M_3}  u_2
                              \end{array} \right) 
  \label{nu2}
 \end{eqnarray}

  where $m^{(2)}_{L^\alpha N_R}$, $M^{(2)}_{N_L N_R}$ and $m^{(2)}_{N_L \nu_R}$ are 
the $2 \times 2$ mass matrices obtained respectively from 
$\mathcal{L}^{(2)}_{L^\alpha N_R}$, $\mathcal{L}^{(2)}_{N_L N_R}$ and 
$\mathcal{L}^{(2)}_{N_L \nu_R}$ terms of (\ref{cly2}). This light neutrino mass 
matrix can be further diagonalized by the bi-unitary transformation. The 
neutrino masses and the mixing angles  obtained from (\ref{nu2}) will be 
dependent on the specific values of the coupling constants $f_i, g_i, M_i$ 
  as well as the VEVs $v, u_i$; $i = 2,3$. For sake of illustration we 
explicitly compute them for two simplifying scenario leading to maximal 
mixing.\\
  
  \textbf{Case I:}  \textbf{$f_2 = f_3 =f$, $g_2 = g_3 = g$, $M_2 = M_3 = M$, 
$u_2 \neq u_3$}. 
  
  In this case the neutrino mass matrix becomes
  
  \begin{eqnarray}
   \mathcal{M}^I_\nu =   \frac{ f g v^*}{M} \, \left( \begin{array}{c c} 
                        u_3        & -  u_2      \\ 
                        u_3        &    u_2
                      \end{array} \right) 
  \label{nu21}
 \end{eqnarray}
 
 \textbf{Case II:}  \textbf{$f_2 = f_3 =f$, $u_2 = u_3 = u$, $\frac{g_2}{M_2} 
\neq \frac{g_3}{M_3}$ }. 
  
  In this case the neutrino mass matrix becomes
  
  \begin{eqnarray}
   \mathcal{M}^{II}_\nu =    f u v^* \, \left( \begin{array}{c c} 
                        \frac{g_2}{M_2}        &  -\frac{g_3}{M_3}      \\ 
                        \frac{g_2}{M_2}        &    \frac{g_3}{M_3}
                      \end{array} \right) 
  \label{nu22}
 \end{eqnarray}
 
  Both the mass matrices in (\ref{nu21}) and (\ref{nu22}) can be written as 
  
 \begin{eqnarray}
   \mathcal{M}_\nu =    \kappa \, \left( \begin{array}{c c} 
                        a        &  -b      \\ 
                        a        &   b
                      \end{array} \right) 
  \label{nu2k}
 \end{eqnarray} 
 
 where
 
 \begin{eqnarray}
  \kappa  \, = \, \left\{ \begin{array}{c }
    \frac{ f g v^*}{M}    \\
    f u v^*             
  \end{array}, \right.
   \quad 
  a  \, = \, \left\{ \begin{array}{c }
     u_3                \\
    \frac{g_2}{M_2}      
  \end{array}, \right. \quad
     b  \, = \, \left\{ \begin{array}{c }
    u_2                 \quad  \quad \rm{In \, Case \, I }   \\
    \frac{g_3}{M_3}       \quad        \quad \rm{In \, Case \, II } 
  \end{array} \right. 
  \label{kvalue}
 \end{eqnarray} 
 
 The mass matrix in (\ref{nu2k}) can be easily diagonalized by a bi-unitary 
transformation leading to the neutrino masses $\nu_2 = \sqrt{2} \kappa a$ 
and $\nu_3= \sqrt{2} \kappa b$ with mixing angle $\theta_{23} = \frac{\pi}{4}$ 
where $\kappa, a, b$ 
 for both cases are given in (\ref{kvalue}). Therefore, $S_3$ allows us to 
understand maximal mixing in the $\mu-\tau$ sector.


\section{The complete $S_3$ invariant lepton sector}
\label{sec3}


 The $B-L$ charge and $S_3$ assignment of the fields for the full lepton 
sector is as shown in Table \ref{tab2}.   

\begin{table}[h]
\begin{center}
\begin{tabular}{c c c || c c c}
  \hline \hline
  Fields                      \hspace{2cm}                    & $B-L$                 \hspace{2cm}                                     &  $S_3$       \hspace{2cm}              & 
  Fields                      \hspace{2cm}                    & $B-L$                 \hspace{2cm}                                     &  $S_3$                                 \\
  \hline \hline
  $L^e$                       \hspace{2cm}                    & $-1$                   \hspace{2cm}                                    &  $1'$            \hspace{2cm}           &   
  $e_R$                       \hspace{2cm}                    & $-1$                    \hspace{2cm}                                   &  $1'$                                    \\
  $L^\mu$                     \hspace{2cm}                    & $-1$                   \hspace{2cm}                                    &  $1'$            \hspace{2cm}           &
  $\mu_R$                     \hspace{2cm}                    & $-1$                   \hspace{2cm}                                    &  $1'$                                    \\
  $L^\tau$                    \hspace{2cm}                    & $-1$                    \hspace{2cm}                                   &  $1$          \hspace{2cm}           &   
  $\tau_R$                    \hspace{2cm}                    & $-1$                    \hspace{2cm}                                   &  $1$                                    \\
  $N^1_L $                    \hspace{2cm}                    & $-1$                    \hspace{2cm}                                   &  $1'$               \hspace{2cm}            &
  $N^1_R $                      \hspace{2cm}                  & $-1$                     \hspace{2cm}                                  &  $1'$                                   \\ 
  $N^2_L $                     \hspace{2cm}                   & $-1$                     \hspace{2cm}                                  &  $1'$               \hspace{2cm}        & 
  $N^2_R $                      \hspace{2cm}                  & $-1$                     \hspace{2cm}                                  &  $1'$                                   \\
  $N^3_L $                      \hspace{2cm}                  & $-1$                    \hspace{2cm}                                   &  $1$                   \hspace{2cm}           & 
  $N^3_R $                        \hspace{2cm}                & $-1$                     \hspace{2cm}                                  &  $1$                                     \\
   $\Phi$                       \hspace{2cm}                  & $0$                      \hspace{2cm}                                  & $1$           \hspace{2cm}               &
  $\nu^e_R $                   \hspace{2cm}                   & $5$                      \hspace{2cm}                                  &  $1'$                                    \\  
       $ \left( \begin{array}{c}
  \nu^\mu_R   \\ \nu^\tau_R   
  \end{array} \right)$               \hspace{2cm}               & $ -4$                     \hspace{2cm}                                 &  $2$            \hspace{2cm}           &    
  \vspace{0.5cm}
  $ \left( \begin{array}{c}
  \chi_2   \\ \chi_3                                                           
  \end{array} \right)$               \hspace{2cm}               & $ 3$                         \hspace{2cm}                             &  $2$                                    \\
  \hline
  \end{tabular}
\end{center}
\caption{The $B-L$ and $S_3$ charge assignment for the fields.}
  \label{tab2}
\end{table}    

The $S_3$ and $B-L$ invariant Yukawa $\mathcal{L}_Y$ interaction can be 
written as 

\begin{eqnarray}
  \mathcal{L}_Y & = &    \mathcal{L}_{L^\alpha l_R} }  \, + \,   
\mathcal{L}_{L^\alpha N_R}   \, + \,  \mathcal{L}_{N_L N_R}      \, + \,  
\mathcal{L}_{N_L \nu_R
 \label{yuk}
\end{eqnarray}

 where 
 
\begin{eqnarray}
\mathcal{L}_{L^\alpha l_R}  & = &  y'_e \, \bar{L}^e \, \Phi \, e_R    \, 
+ \, y'_{12} \, \bar{L}^e \, \Phi \, \mu_R    \, + \, y'_{21} \, \bar{L}^\mu \, 
\Phi \, e_R    \, + \, y'_\mu \, \bar{L}^\mu \, \Phi \, \mu_R    
\, + \,   y_\tau \,  \bar{L}^\tau \, \Phi \, \tau_R        \nonumber \\
\mathcal{L}_{L^\alpha N_R}  & = &  g'_{11} \, \bar{L}^e \, \hat{\Phi}^* \, N^1_R    
\, + \, g'_{12} \, \bar{L}^e \, \hat{\Phi}^* \, N^2_R    \, + \, g'_{21} \, 
\bar{L}^\mu \, \hat{\Phi}^* \, N^1_R    \, + \,g'_{22} \, \bar{L}^\mu \, 
\hat{\Phi}^* \, N^2_R  
\, + \,  g_{33}  \, \bar{L}^\tau \, \hat{\Phi}^* \, N^3_R     \nonumber \\
\mathcal{L}_{N_L N_R}  & = & M'_{11} \, \bar{N}^1_L \, N^1_R  \, + \, M'_{12} \, 
\bar{N}^1_L \, N^2_R  \, + \, M'_{21} \, \bar{N}^2_L \, N^1_R  \, + \,  M'_{22} \, 
\bar{N}^2_L \, N^2_R  \, + \,  M_{33} \, \bar{N}^3_L  \, N^3_R   \nonumber \\
\mathcal{L}_{N_L \nu_R}  & = &   \frac{f'_{11}}{\Lambda} \, \left (\bar{N}^1_L \, 
\nu^e_R \right) \otimes \left[ \left( \begin{array}{c}  \chi^*_3   \\ 
\chi^*_2   \end{array} \right) \, \otimes \, \left( \begin{array}{c}  
\chi^*_3   \\ \chi^*_2  \end{array} \right) \right]_{1} 
\; + \; \frac{f'_{21}}{\Lambda} \, \left (\bar{N}^2_L \, \nu^e_R \right) 
\otimes \left[ \left( \begin{array}{c}  \chi^*_3   \\ \chi^*_2   \end{array} 
\right) \, \otimes \, \left( \begin{array}{c}  \chi^*_3   \\ \chi^*_2  
\end{array} \right) \right]_{1} 
\nonumber \\
& + &  f'_{12} \, \bar{N}^1_L \, \otimes \, \left[  \left( \begin{array}{c}  
\nu^\mu_R   \\ \nu^\tau_R  \end{array} \right) \, \otimes \, \left( 
\begin{array}{c}  \chi_2   \\ \chi_3   \end{array} \right) \right]_{1'}
\; + \; f'_{22} \, \bar{N}^2_L \, \otimes \, \left[  \left( \begin{array}{c}  
\nu^\mu_R   \\ \nu^\tau_R  \end{array} \right) \, \otimes \, \left( 
\begin{array}{c}  \chi_2   \\ \chi_3   \end{array} \right) \right]_{1'} 
\nonumber \\
& + &  f_{33} \, \bar{N}^3_L \, \otimes \, \left[  \left( \begin{array}{c}  
\nu^\mu_R   \\ \nu^\tau_R  \end{array} \right) \, \otimes \, \left( 
\begin{array}{c}  \chi_2   \\ \chi_3   \end{array} \right) \right]_{1}  
\label{cly}
\end{eqnarray}

As before, in writing (\ref{cly}), we used the notation $\hat{\Phi}^* = 
i \tau_2 \Phi^* = (\phi^0, \phi^-)^{\rm{T}}$. Moreover, $y_\alpha$ are the 
Yukawa couplings of the charged leptons whereas  $f_{ij}$, $g_{ij}$ and $M_{ij}$ 
denote the dimensionless coupling constants between the leptons and the 
heavy fermions.      \\

At this point we like to remark that in (\ref{cly}) there is still a 
freedom to redefine a few fields (i.e. the pairs $N^1_{L} - N^2_L$, 
$N^1_R - N^2_R$ and $e_R - \mu_R$) in a way that certain couplings can be 
made equal to zero.  For sake of later convenience
we choose to use this freedom of field redefinition to make 
$f'_{11} = M'_{12} = y'_{21} =0$. Moreover, we relabel the remaining 
non-zero couplings of these redefined fields as $f'_{ij} \to f_{ij}, g'_{ij} 
\to g_{ij}, M'_{ij} \to M_{ij}$. \\ 

Now, after symmetry breaking the scalar fields get VEVs $\langle \phi^0 
\rangle = v$, $\langle \chi_i \rangle = u_i$; $i = 2,3$.  Then the mass 
matrices relevant to charged lepton is given by 

 \begin{eqnarray}
   \mathcal{M}_l =   v \, \left( \begin{array}{c c c} 
                      y_e     &  y_{12}    &  0     \\
                      0         &  y_\mu       &  0      \\ 
                      0         &  0          &   y_\tau
                      \end{array} \right) 
  \label{lm}
 \end{eqnarray}
 
The mass matrix (\ref{lm}) can be readily diagonalized by bi-unitary 
transformation. In the limit of $y_e << y_\mu$ we get 

  \begin{eqnarray}
    \theta^l_{12} & \approx & \tan^{-1}\left( \frac{-y_{12}}{y_\mu} \right); 
\qquad \qquad \quad  m_e \approx v \, y_e \rm{cos} \, \theta^l_{12}   
\nonumber \\
    m_\mu & \approx & v \left( y_\mu \rm{cos} \, \theta^l_{12} - y_{12} 
\rm{sin}\, \theta^l_{12}   \right) ; \qquad m_\tau \approx v y_\tau
   \label{lmu}
  \end{eqnarray}
 
 If  $y_{12} = y_\mu$ then maximal mixing is achieved i.e. 
$ \theta^l_{12} = -\frac{\pi}{4}$, with $m_\mu = \sqrt{2} v y_\mu$.     \\

Also, the $6 \times 6$ mass matrix spanning $(\bar{\nu}^e_L, \bar{\nu}^\mu_L, 
\bar{\nu}^\tau_L, \bar{N}^1_L, \bar{N}^2_L, \bar{N}^3_L)$ and $(\nu^e_R, 
\nu^\mu_R, \nu^\tau_R, N^1_R, N^2_R, N^3_R)^{\rm{T}}$ of neutrinos and the 
heavy fermions is given by

\begin{eqnarray}
   \mathcal{M}_{\nu,N} =    \, \left( \begin{array}{c c  c c c c} 
                      0                           &   0                  & 0                  &   g_{11} v^*        &    g_{12} v^*             &   0                       \\ 
                      0                           &   0                  &  0                 &   g_{21} v^*        &    g_{22}  v^*            &   0                        \\ 
                      0                           &   0                  &  0                 &   0                 &    0                      &   g_{33}  v^*              \\ 
                      0                           &   f_{12} u_3         & -f_{12} u_2        &  M_{11}             &    0                      &   0                         \\ 
         \frac{f_{21}}{\Lambda} u^*_2 u^*_3       &   f_{22} u_3         & -f_{22} u_2        &  M_{21}             &    M_{22}                 &   0                         \\ 
                      0                           &   f_{33} u_3         &   f_{33} u_2       &  0                  &    0                      &  M_3 
                      \end{array} \right) 
  \label{lnuN}
 \end{eqnarray}

 As remarked earlier, the mass terms $M_{ij}$ between the heavy fermions can be 
naturally large, so we can block diagonalize (\ref{lnuN}) assuming that 
$f_{ij}, g_{ij} << M_{ij}$. The block diagonalized mass matrix of light 
neutrinos is given by
 
 \begin{eqnarray}
 \mathcal{M}_\nu  & = &   m_{N_L \nu_R} \frac{1}{M_{N_L N_R}} m_{L^\alpha N_R}  \nonumber \\
 & = &  v^*  \left( \begin{array}{c c c} 
 \frac{(g_{21} M_{11} - g_{11} M_{21} )f_{12} u_2}{M_{11} M_{22}}                                                                &        \frac{(g_{22} M_{11} - g_{12} M_{21}) f_{12} u_2}{M_{11} M_{22}}               
            &             \frac{-f_{12} g_{33} u_3}{M_{33}}                  \\
 \frac{(g_{21} M_{11} - g_{11} M_{21} )f_{22} u_2 \, +\, f_{21} g_{11} M_{22} u_6}{M_{11} M_{22}}        &      \frac{(g_{22} M_{11} - g_{12} M_{21}) f_{22} u_2 \, +\, f_{21} g_{12} M_{22} u_6}{M_{11} M_{22}}    
            &            \frac{-f_{22} g_{33} u_3}{M_{33}}     \\ 
\frac{(g_{21} M_{11} - g_{11} M_{21} )f_{33} u_2}{M_{11} M_{22}}                                                                 &       \frac{(g_{22} M_{11} - g_{12} M_{21}) f_{33} u_2}{M_{11} M_{22}}        
            &             \frac{f_{33} g_{33} u_3}{M_{33}} 
                      \end{array} \right) 
  \label{nu}
 \end{eqnarray}

where we have written $u_6 = \frac{u^*_2 u^*_3}{\Lambda}$.  Also, the $3 \times 
3$ mass matrices $m_{L^\alpha N_R}$, $M_{N_L N_R}$ and $m_{N_L \nu_R}$ are obtained 
from the terms $\mathcal{L}_{L^\alpha N_R}$, $\mathcal{L}_{N_L N_R}$ and 
$\mathcal{L}_{N_L \nu_R}$
of (\ref{cly}), respectively. This light neutrino mass matrix can be further 
diagonalized by the bi-unitary transformation. The neutrino masses and the 
mixing angles  obtained from (\ref{nu}) will be dependent on the specific 
values of the coupling constants $f_{ij}, g_{ij}, M_{ij}$  as well as the 
VEVs $v, u_i$; $i = 2,3$. \\
  
In the simplifying case of $g_{ij} = g$ and $M_{ij} = M$ we get 
  
\begin{eqnarray}
 \mathcal{M}_\nu =   \frac{g v^*}{M}  \left( \begin{array}{c c c} 
               0             &         0                       &             -f_{12}  u_3                  \\
          f_{21}  u_6        &     f_{21}  u_6                 &            -f_{22}  u_3     \\ 
               0             &         0                       &             f_{33} u_3
                      \end{array} \right) 
  \label{nus}
\end{eqnarray}
 
Diagonalizing the mass matrix (\ref{nus}) we have 

  \begin{eqnarray}
    \theta^\nu_{12} & \approx & 0 ;  \quad \quad  \theta^\nu_{13}  \approx   \tan^{-1}\left( \frac{f_{12}}{f_{33}}  \right);     \quad  \quad   \theta^\nu_{23}  \approx   \tan^{-1}\left( \frac{f_{22}}{\sqrt{f^2_{12} + f^2_{33}}}  \right)    
    \nonumber \\
   m^\nu_1 & \approx & 0 ; \quad \quad  m^\nu_2  \approx  \frac{\sqrt{2(f^2_{12} + f^2_{33})} f_{21} g |v|}{M \sqrt{f^2_{12} + f^2_{22} + f^2_{33}}}\, |u_6|;   \quad \quad    
   m^\nu_3  \approx    \frac{\sqrt{f^2_{12} + f^2_{22} + f^2_{33}}\, g |v|}{M }\, |u_3|
   \label{numu}
  \end{eqnarray}
 
 Since, $u_6 << u_3$, we have a normal hierarchy pattern with two nearly 
massless neutrinos and one relatively heavy neutrino. Moreover, the massless 
neutrino will also gain small mass,  if any of the $M_{ij}$'s or $g_{ij}$'s 
are not equal to $M$ or $g$ 
 respectively. Also, if they deviate significantly from these values then one 
can possibly recover degenerate or inverted hierarchy patterns also.  \\
 
Now, if $U_{l}$ and $U_\nu$ are the mixing matrices of the charged leptons 
and neutrinos respectively, then the PMNS mixing matrix is given by 

\begin{eqnarray}
  U_{\rm{PMNS}}  = U^\dagger_{l} U_\nu
 \label{pmns}
\end{eqnarray}

Taking $y_{12} = y_\mu$, $f_{12} = -\frac{f_{33}}{2}$ and $f_{22} = \sqrt{f^2_{12} 
+ f^2_{33}}$ in (\ref{lmu}) and (\ref{numu}) we get $\theta^\nu_{23} 
= -\theta^l_{12} =  \frac{\pi}{4}$ and $\theta^\nu_{13} = 
\tan^{-1}(-\frac{1}{2})$ which gives PMNS mixing angles
 consistent\footnote{With $\theta^\nu_{23} = -\theta^l_{12} =  
\frac{\pi}{4}$ and $\theta^\nu_{13} = \tan^{-1}(-\frac{1}{2})$ we get the 
value of mixing angles as $\theta_{12} = 30.29^\circ$, $\theta_{13} = 
7.53^\circ$ and $\theta_{23} = 50.36^\circ$. The values of $\theta_{12}$ 
and $\theta_{13}$ are slightly below the lower limits ($30.59^\circ$, 
$7.62^\circ$ respectively) quoted in \cite{Capozzi:2013csa}. A much better 
fit can be obtained if we take slightly different values, for instance 
taking $\theta^l_{12} = -50^\circ$, $\theta^\nu_{13} = -29.39^\circ$ and  
$\theta^\nu_{23} =  \frac{\pi}{4}$ gives us $\theta_{12} = 33.26^\circ$, 
$\theta_{13} = 9.00^\circ$ and $\theta_{23} = 51.41^\circ$. } with present 
$3-\sigma$ limits of global fits obtained from 
experiments \cite{Capozzi:2013csa}.

 
\subsection*{The Quark Sector }
\label{qsec} 
 
  
In our minimal model with only one doublet scalar, the quark sector can be 
accommodated in a simple way if both the left handed quark doublets 
$Q^i_L = (u^i_L, d^i_L)^{\rm{T}}$, $i= 1,2,3$ and the right handed quark 
singlets $u^i_R$, $d^i_R$; $i=1,2,3$
transform as $1$ of $S_3$. \\

A better understanding of the quark sector can be obtained if, to our 
minimal model, we add more doublet scalars transforming non-trivially 
under $S_3$. One such example for quark sector, albeit in context of a 
different model for lepton sector, has already been worked out 
in \cite{Chen:2004rr, Ma:2013zca}. We plan to present a similar extension 
of our minimal model in a future work.


\section{Conclusions}
\label{sec6}

The idea that $B-L$ should be a gauge symmetry has been around for a long 
time.  The minimal version of three $\nu_R$ transforming as $-1$ is very 
well-known.  Other exotic variants are possible with extra particles, such 
as the two models recently proposed \cite{Kanemura:2011mw,Kanemura:2014rpa}.  Here we point out 
the simple anomaly-free solution of three $\nu_R$'s transforming as 
$+5,-4,-4$.  We show how these assignments may be used to obtain seesaw 
Dirac neutrino masses, as well as 
inverse seesaw Majorana neutrino masses.  We then apply $S_3$ symmetry 
to the first case, and obtain realistic neutrino and charged-lepton 
mass matrices with a mixing pattern consistent with experiments.


\begin{acknowledgments}
RS will like to thank G. Rajasekaran for valuable comments and suggestions. EM is supported in part by the U. S. Department of Energy under 
Grant No. DE-SC0008541.
\end{acknowledgments}



\begin{thebibliography}{99}



\bibitem{Ma:2001kg} 
  E.~Ma,
  Mod.\ Phys.\ Lett.\ A {\bf 17}, 535 (2002)
  [hep-ph/0112232].

\bibitem{Ma:2002tc} 
  E.~Ma,
  Phys.\ Rev.\ Lett.\  {\bf 89}, 041801 (2002)
  [hep-ph/0201083].
  
\bibitem{Montero:2007cd} 
  J.~C.~Montero and V.~Pleitez,
  Phys.\ Lett.\ B {\bf 675}, 64 (2009)
  [arXiv:0706.0473 [hep-ph]].

  
\bibitem{Machado:2010ui} 
  A.~C.~B.~Machado and V.~Pleitez,
  Phys.\ Lett.\ B {\bf 698}, 128 (2011)
  [arXiv:1008.4572 [hep-ph]].
  
\bibitem{Machado:2013oza} 
  A.~C.~B.~Machado and V.~Pleitez,
  J.\ Phys.\ G {\bf 40}, 035002 (2013)
  [arXiv:1105.6064 [hep-ph]].

  
\bibitem{Roy:1983be} 
  P.~Roy and O.~U.~Shanker,
  Phys.\ Rev.\ Lett.\  {\bf 52}, 713 (1984)
  [Erratum-ibid.\  {\bf 52}, 2190 (1984)].
  
  
\bibitem{Wyler:1982dd} 
  D.~Wyler and L.~Wolfenstein,
  Nucl.\ Phys.\ B {\bf 218}, 205 (1983).

\bibitem{Mohapatra:1986bd} 
  R.~N.~Mohapatra and J.~W.~F.~Valle,
  Phys.\ Rev.\ D {\bf 34}, 1642 (1986).
 
\bibitem{Ma:1987zm} 
  E.~Ma,
  Phys.\ Lett.\ B {\bf 191}, 287 (1987).
  
  
\bibitem{Ma:2009du} 
  E.~Ma,
  Mod.\ Phys.\ Lett.\ A {\bf 24}, 2161 (2009)
  [arXiv:0904.1580 [hep-ph]].
 
\bibitem{Chen:2004rr} 
  S.~L.~Chen, M.~Frigerio and E.~Ma,
  Phys.\ Rev.\ D {\bf 70}, 073008 (2004)
  [Erratum-ibid.\ D {\bf 70}, 079905 (2004)]
  [hep-ph/0404084].
 
 
  
\bibitem{Ma:2013zca} 
  E.~Ma and B.~Melic,
  Phys.\ Lett.\ B {\bf 725}, 402 (2013)
  [arXiv:1303.6928 [hep-ph]].
 
\bibitem{Capozzi:2013csa} 
  F.~Capozzi, G.~L.~Fogli, E.~Lisi, A.~Marrone, D.~Montanino and A.~Palazzo,
  Phys.\ Rev.\ D {\bf 89}, 093018 (2014)
  [arXiv:1312.2878 [hep-ph]].
  
\bibitem{Kanemura:2011mw} 
  S.~Kanemura, T.~Nabeshima and H.~Sugiyama,
  Phys.\ Rev.\ D {\bf 85}, 033004 (2012)
  [arXiv:1111.0599 [hep-ph]].

\bibitem{Kanemura:2014rpa} 
  S.~Kanemura, T.~Matsui and H.~Sugiyama,
  Phys.\ Rev.\ D {\bf 90}, 013001 (2014)
  [arXiv:1405.1935 [hep-ph]].

  
  

\end{thebibliography}
\end{document}